\documentclass[seceq,amsmath,amssymb]{ptptex}

\usepackage{graphicx}
\usepackage{bm}

\def\bea{\begin{eqnarray}}
\def\eea{\end{eqnarray}}

\def\pp{\mbox{$p$-$p$} }
\def\auau{\mbox{Au-Au} }
\def\cucu{\mbox{Cu-Cu} }

\def\aa{\mbox{A-A} }

\title{
Review of soft interactions and multiparticle correlations%
}

\author{
Thomas A. \textsc{Trainor}%
}

\inst{
CENPA 354290, University of Washington, Seattle, WA 98195 USA
}

\abst{
Recent developments in the area of soft interactions and multiparticle correlations are reviewed, including higher harmonic flows conjectured to result from the initial-state geometry which might coincide with some jet-like correlation features in more-central \aa collisions.
}

\begin{document} 
\maketitle

 \section{Introduction}

The subject of this review has seen important new developments in the past year. Observed deviations in more-central \aa collisions from perturbative QCD (pQCD) expectations for angular correlations from jets include (a) elongation on $\eta$ of the same-side (SS) 2D peak from a nominally-symmetric 2D jet cone to what is called by some a ``soft ridge''~\cite{axialci,anomalous} and (b) distortion of the away-side (AS) 1D peak on $\phi$ representing back-to-back jet pairs to a double-peaked structure interpreted by some as evidence for ``Mach cones.''~\cite{tzyam} Both features have been reinterpreted recently as manifestations of ``higher harmonic flows,'' including ``triangular flow,''~\cite{gunther,multipoles} arising from initial-state (IS) geometry fluctuations~\cite{sorensen} or glasma flux tubes.~\cite{gmm,glasma}
Those developments are part of an ongoing competition between jets and flows to describe the hadronic final state of nuclear collisions. The conventional jet scenario is replaced by conjectured features of the IS geometry coupled with radial flow to produce final-state (FS) flow structures which may coincide with some jet-like correlation features.

\section{Flows vs jets: What mechanism dominates \aa collisions?}

The ongoing competition between jets and flows for $p_t$ ``real estate'' has emerged as a dramatic struggle for possession of the \aa hadronic final state. At lower $p_t$ flows (especially elliptic flow) are assumed to dominate. At higher $p_t$ jets should dominate. The location of the jet-flow boundary on $p_t$ is all-important since particle density falls rapidly with $p_t$. The minimum-bias SS 2D peak occupies a strategic intermediate position at $p_t \approx 1$ GeV/c (parton energy scale $Q \approx 6$ GeV). Proponents of higher harmonic flows wish to interpret the SS 2D peak in terms of IS structure (glasma flux tubes, geometry fluctuations) coupled with radial flow. However, there is substantial experimental and theoretical evidence to support a jet interpretation.

The role of jets in nuclear collisions may be underestimated if established jet systematics are not fully appreciated. The properties of flows, even their existence, may then be misunderstood. Initial-state geometry conjectures as expressed in Monte Carlos may produce final-state (FS) structure similar {\em in some respects}  to the observed final state. But other aspects of correlation data may contradict such conjectures. The $p_t$ and $\eta$ dependence of 2D angular correlations play key roles. Alternative descriptions should be confronted by all features of 2D angular correlations.

 \section{Autocorrelations, impulse responses and transfer functions}

Jets and flows share a dual description system (LTI or linear time-invariant system theory).  LTI theory was originally developed in terms of time and frequency domains to describe stochastic time evolution. For application to nuclear collisions corresponding space and wave-number domains are appropriate. The {\em autocorrelation} formalism was developed to describe stochastic elements of time evolution (e.g., Langevin equation) but is also appropriate for nuclear collision data. LTI theory predicts the output of a process for a given input based on a fixed property of the process: the impulse response (space or time domain) or transfer function (wave-number or frequency domain). The input, output and process property in the dual domains are related by the Fourier transform (more generally the Laplace transform).

 \begin{figure}[h]
 \hfill \includegraphics[width=1.65in,height=1.4in]{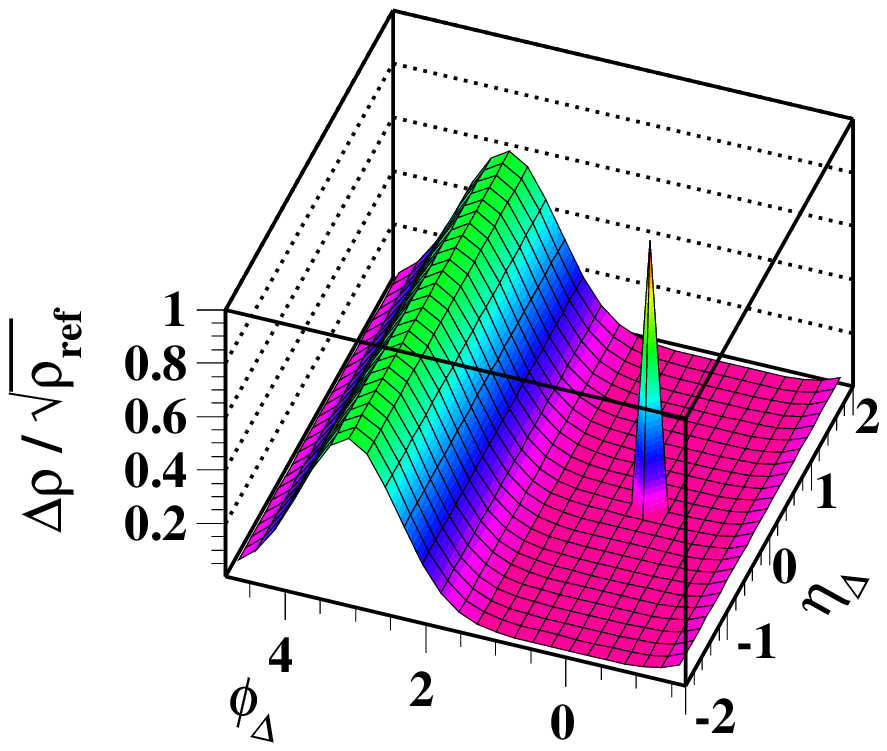}
  \includegraphics[width=1.65in,height=1.4in]{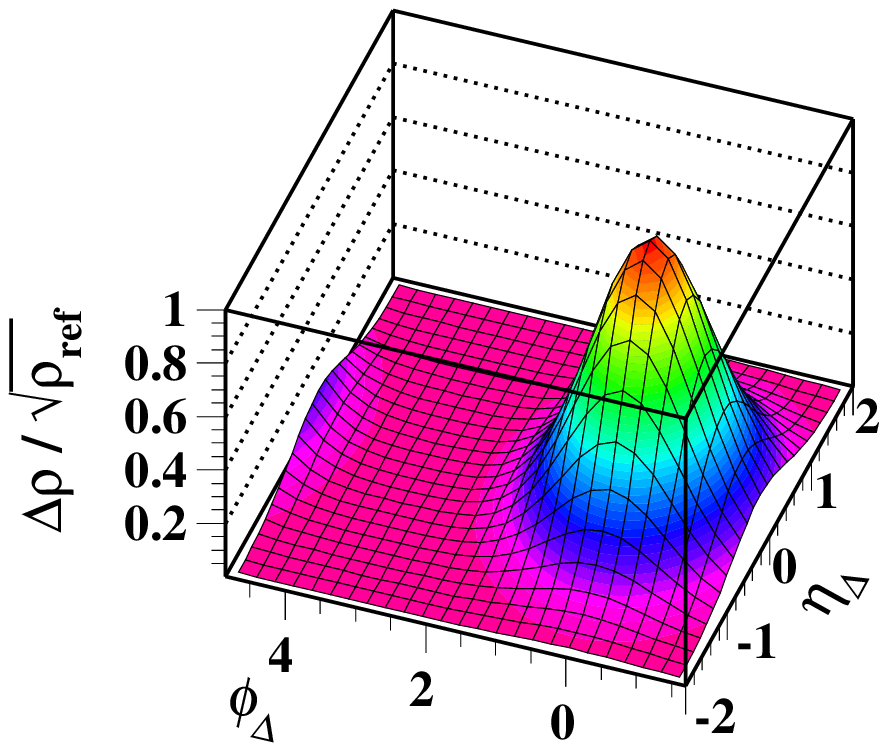}
  \includegraphics[width=1.65in,height=1.4in]{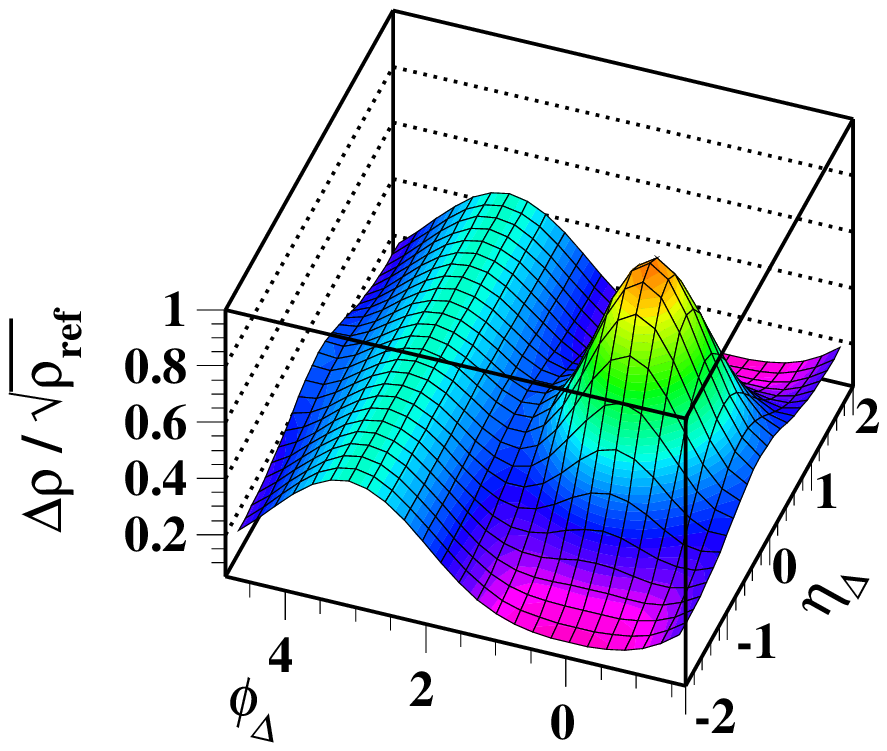}
\hfill
\hfill
\caption{\label{autocorr}
First: Input autocorrelation,
Second: Impulse response,
Third: Output autocorrelation.
}
 \end{figure}

Figure~\ref{autocorr} illustrates LTI theory applied to jets. The first panel is the input parton momentum-space angular autocorrelation, with a self-pair peak at the origin and an AS peak at $\pi$ representing back-to-back parton pairs, with acoplanarity (broadening) from IS $k_t$. The second panel is the impulse response (fragmentation function angular correlation) for parton fragmentation. The third panel is the output hadron angular correlation obtained by folding the input with the impulse response. The folding integral is the QCD ``factorization theorem.'' The result describes \pp data.~\cite{porter}

 \begin{figure}[h]
 \hfill \includegraphics[width=1.65in,height=1.4in]{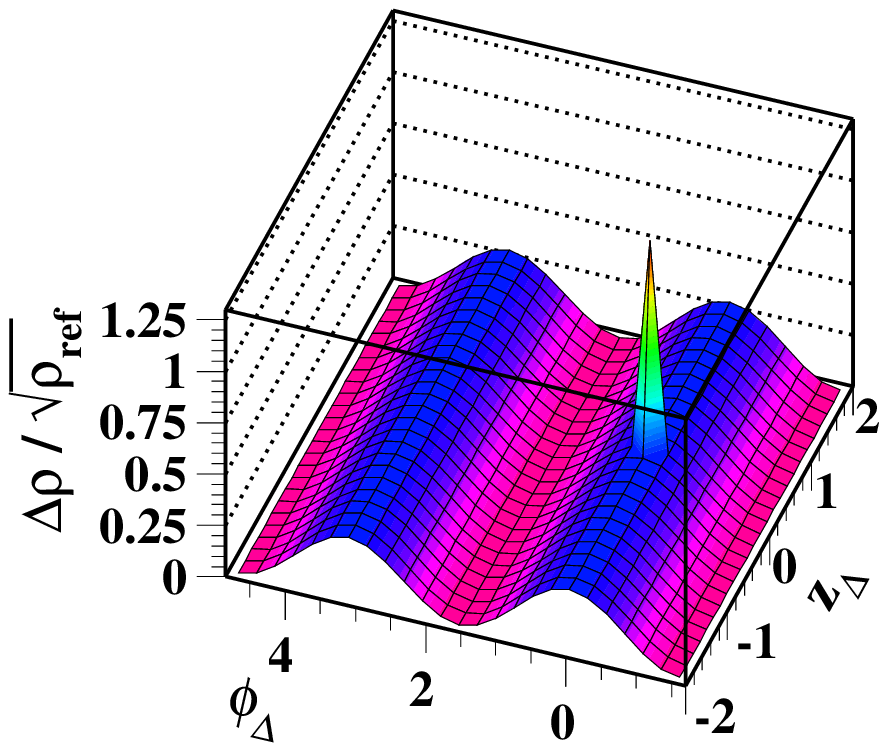}
  \includegraphics[width=1.65in,height=1.4in]{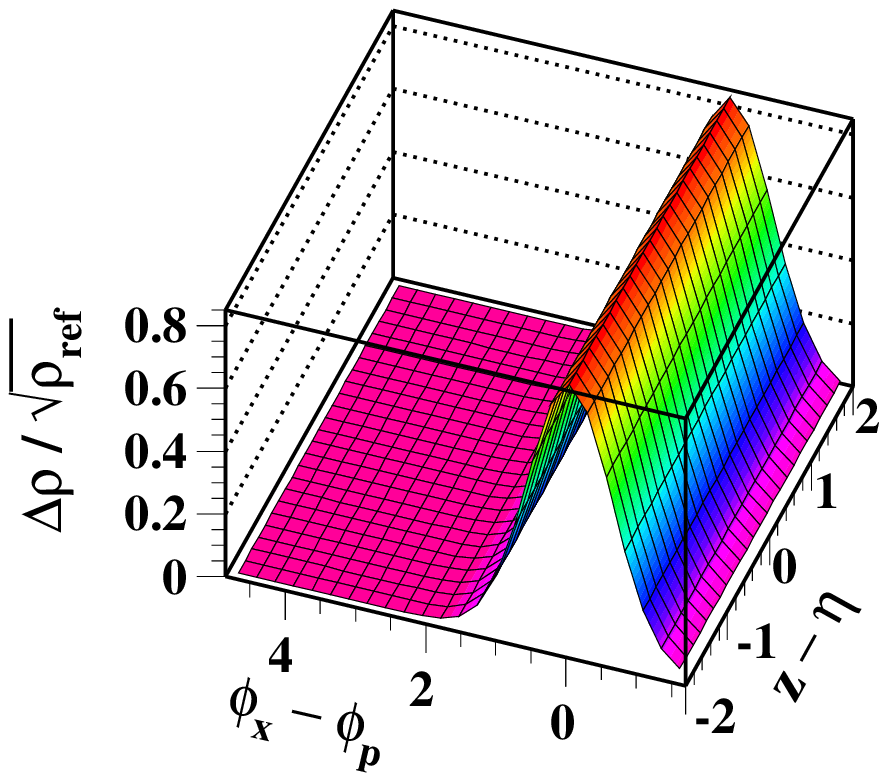}
  \includegraphics[width=1.65in,height=1.4in]{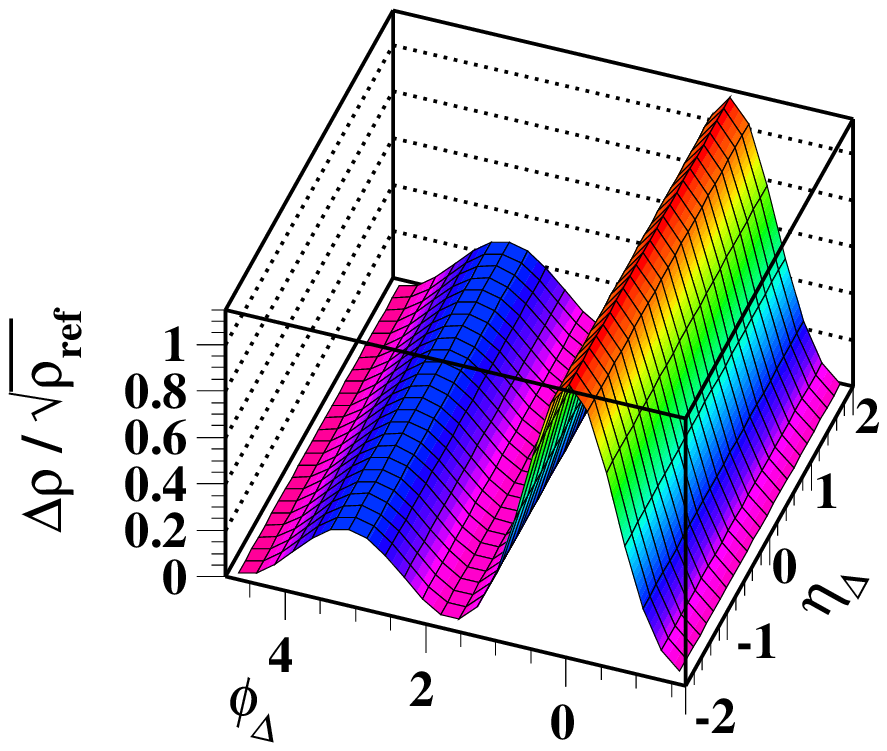}
\hfill
\hfill
\caption{\label{room}
First: Input autocorrelation,
Second: Impulse response,
Third: Output autocorrelation.
}
 \end{figure}

Figure~\ref{room} illustrates LTI theory applied to flows, {\em also in the space domain}. The first panel is a configuration-space angular autocorrelation representing the geometry of noncentral \aa collisions (sinusoids) and a self-pair peak arising from point sampling in a Monte Carlo. The second panel illustrates an impulse response for radial flow. The third panel is obtained by folding the input with the impulse response.

Jets in elementary collisions are more efficiently represented as localized (on $\eta$) structures in the momentum-space domain, whereas conjectured flows are more efficiently represented as uniform (on $\eta$) azimuth sinusoids in the wave-number domain, with the conjectured Gaussian radial-flow impulse response on azimuth transformed to a transfer function (also Gaussian) on wave number. However, the alternative dual representations in each case (jets or flows) are equally valid albeit less efficient. 

Space and wave-number domains are typically interpreted in terms of jets and flows respectively. When the SS 2D peak becomes elongated (in more-central \aa collisions) the more-efficient representation is ambiguous. If the SS peak is projected onto azimuth and the wave-number domain is elected its physical interpretation can be shifted (perhaps erroneously) to flows. In the wave-number domain conjectured IS geometry multipoles are multiplied by a conjectured collective-expansion transfer function to produce FS ``higher harmonic flow'' multipoles which {\em seem} to match the Fourier components of an elongated SS 2D jet peak {\em if projected onto 1D azimuth}.

 \section{Minimum-bias jets (minijets)}

The terms ``soft'' and ``hard'' apply both to the initial-state (IS) four-momentum transfer between colliding hadrons or their partonic constituents and to the momenta of final-state (FS) hadrons produced in the hadronization process, whether from a fragmentation cascade or from conjectured bulk-medium freezeout. All jets are dominated by low-$p_t$ (soft) fragments ($< 2$ GeV/c).~\cite{ffprd} The observed parton spectrum is consistent with pQCD down to 3 GeV.~\cite{ua1,fragevo} The minimum-bias jet ensemble is thus dominated by 3 GeV jets (minijets). ``Soft interactions'' includes jet structure.

 \begin{figure}[h]
\hfill
  \includegraphics[width=1.65in,height=1.5in]{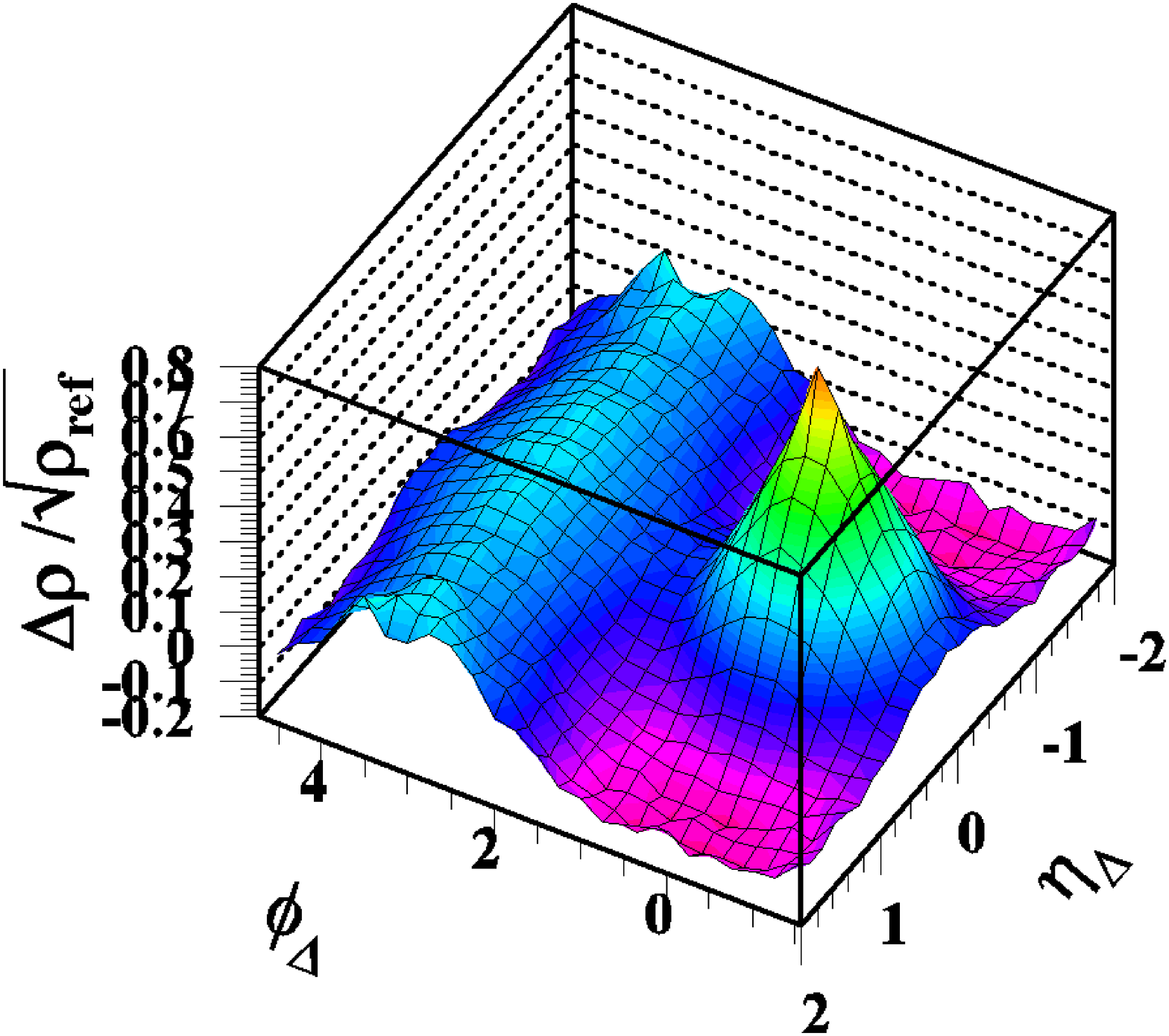}
  \includegraphics[width=1.65in,height=1.5in]{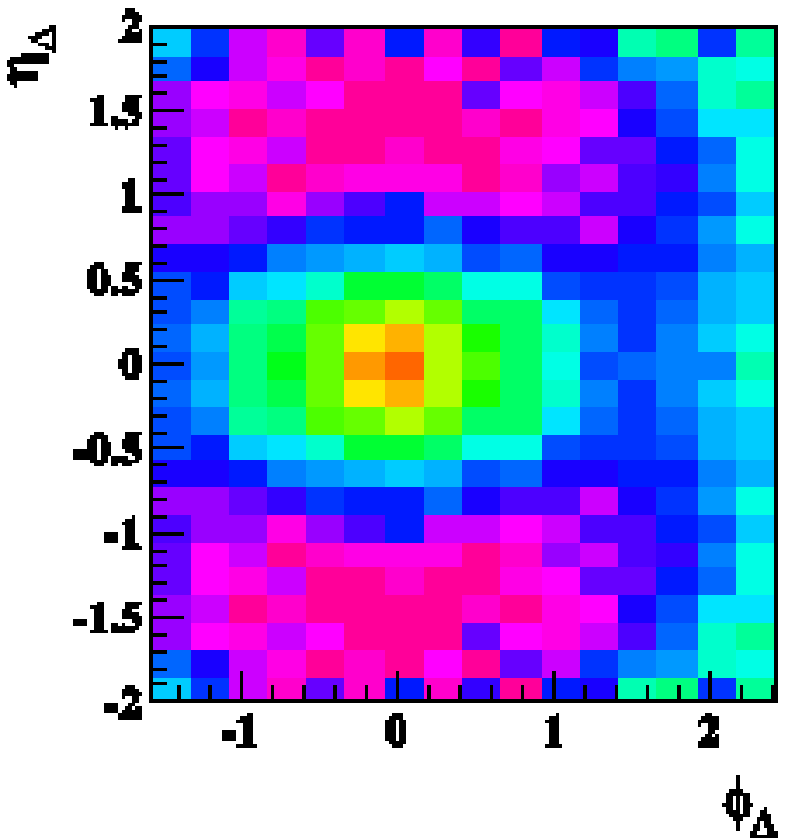}
  \includegraphics[width=1.65in,height=1.5in]{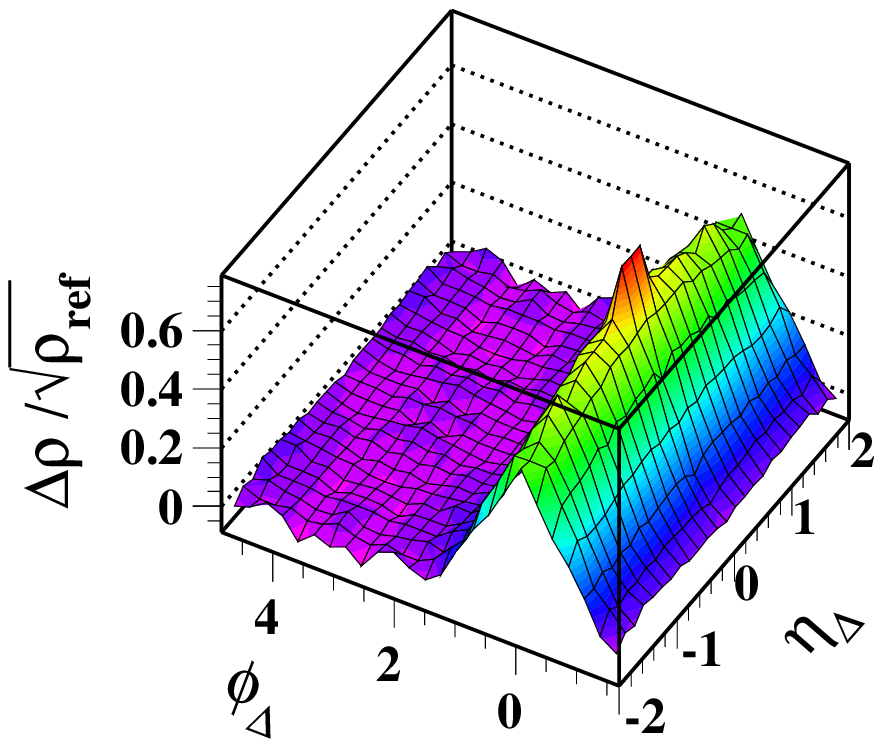}
\hfill
\hfill
\caption{\label{minijets}
First: \pp correlations,
Second: Elongation on $\phi$ in \pp,
Third: SS peak in central Au-Au.
}
 \end{figure}

Figure~\ref{minijets} illustrates some minijet features. The first panel shows pQCD jet-like angular correlations for hadron $p_t \approx 0.6$ GeV/c.~\cite{porter} Minijet phenomenology shows clear jet structure for hadron $p_t$ down to 0.35 GeV/c in \pp collisions. The second panel shows strong elongation (2:1) {\em on azimuth} of the SS 2D peak in \pp collisions. The third panel shows the isolated SS 2D peak for 0-5\% 200 GeV \auau collisions elongated (3:1) on $\eta.$~\cite{anomalous} The AS dipole (back-to-back jets) has been subtracted.

\section{Nonjet azimuth quadrupole}

The nonjet azimuth quadrupole derived from model fits to 2D angular correlations~\cite{anomalous} and measure by $A_Q\{2D\}(b) \equiv \rho_0(b) v_2^2\{2D\}(b)$ exhibits very simple centrality and energy systematics given by~\cite{davidhq} 
\bea
A_Q\{2D\}(b) &=& 0.0045 R(\sqrt{s_{NN}}) N_{bin} \epsilon_{opt}^2
\eea
where $R(\sqrt{s_{NN}}) = \log(\sqrt{s_{NN}} / \text{13.5 GeV}) / \log(200 / 13.5)$ measures the observed energy trend relative to 200 GeV, and $\rho_0(b) = dn_{ch} / 2\pi d\eta$ is the single-particle density. $A_Q\{2D\}(b)$ is statistically compatible with $A_{2D}$ which measures the amplitude of the SS 2D (minijet) peak. The 2D model fit accurately separates jet structure (``nonflow'') from the nonjet quadrupole (``elliptic flow'').~\cite{anomalous} The nonjet quadrupole is a unique phenomenon independent of SS and AS jet structure and does not exhibit the {\em sharp transition} which is a prominent feature of minijet centrality dependence.~\cite{anomalous} Most $v_2$ measurement methods are equivalent to fitting a cosine function $\cos(2\phi)$ to 2D angular correlations projected onto 1D azimuth, in which case some part of the jet structure may be included in the $v_2$ data  as a (nonflow) bias. Various methods are used to estimate the nonflow bias.~\cite{2004} Biased $v_2$ data used to estimate backgrounds in 1D azimuth (dihadron) correlations may lead to distorted inferred jet correlations.~\cite{tzyam}

\section{Summary}

``Soft interactions and multiparticle correlations'' currently addresses a critical conflict between jet and flow interpretations of nuclear collisions centering on properties of the same-side 2D peak in angular correlations. $\eta$ elongation of the SS 2D peak in more-central \aa collisions has lead to interpretation ambiguity. Projection to 1D azimuth seems consistent with higher harmonic flows.  To reduce ambiguity Fourier transformations between dual azimuth and wave-number representations should {\em retain all structure on $\eta$}. The so-called "soft ridge" may be  jets modified (e.g., polarized) by strong longitudinal chromoelectric fields in \aa collisions. 
Resolution of conflicting interpretations of correlation structure requires careful study of analysis methods and phenomenology and may lead to important new QCD physics.


\end{document}